\documentclass{aa}
\usepackage{psfig}

\newcommand{\MC}{\multicolumn}
\newcounter{qub}
\begin{document}

\thesaurus{20(04.19.1; 11.06.2; 11.04.1; 11.19.3; 11.03.2)}
\title{HS 0822+3542 -- a new nearby extremely metal-poor galaxy}

\author{Kniazev A.Y.\inst{1}
\and Pustilnik S.A.\inst{1}
\and Masegosa J.\inst{2}
\and M\'arquez I.\inst{2}
\and Ugryumov A.V.\inst{1}
\and Martin J.-M.\inst{3}
\and \\ Izotov Y.I.\inst{4}
\and  Engels D.\inst{5}
\and Brosch N.\inst{6}
\and Hopp U.\inst{7}
\and Merlino S.\inst{2}
\and Lipovetsky V.A.\inst{1}\thanks{Deceased 1996 September 22.}}

\offprints{A. Kniazev, akn@sao.ru}

\institute{
Special Astrophysical Observatory, Nizhnij Arkhyz, Karachai-Circessia,
369167, Russia
\and Instituto de Astrofisica de Andalucia, CSIC, Aptdo. 3004, 18080
   Granada, Spain
\and Observatoire de Paris-Meudon, 92195 Meudon, France
\and Main Astronomical Observatory, Goloseevo, Kiev-127, 03680, Ukraine
\and Hamburger Sternwarte, Gojenbergswerg 112, D-21029 Hamburg, Germany
\and Wise Observatory, Tel-Aviv University, Tel-Aviv 69978, Israel
\and Universit\"atssternwarte M\"unchen, D-81679 M\"unchen, Germany
}

\date{Received \hskip 2cm; Accepted}

\maketitle

\markboth{A.Kniazev et al.: HS 0822+3542 -- an extremely metal-poor galaxy}
{HS 0822+3542 -- an extremely metal-poor galaxy}

\begin{abstract}
We present the results of spectrophotometry and $BVR$ CCD photometry, 
as well as integrated H{\sc i} radio measurements of a new blue compact galaxy
(BCG) HS~0822+3542 with extremely low oxygen abundance: 12 + log(O/H) = 7.35, 
or 1/36 of solar value. The galaxy is the
third most metal-deficient BCG after I~Zw~18 and SBS~0335--052. Its
very high mass fraction of gas ($\approx$ 95\% of all visible mass)
and blue colours of underlying nebulosity are also similar to those of
SBS~0335--052. This suggests that HS~0822+3542 is one of the nearest and
dimmest galaxies
experiencing a recently-started first star formation (SF) episode. 
Its properties imply that for such galaxies there is a
linear scaling of the main parameters, at least for the baryon mass range
(0.3--20)$\times$10$^{8}$ $M_{\odot}$.
The total mass estimate indicates that the galaxy is
dynamically dominated by a dark matter (DM) halo, which itself is one of
the least massive for galaxies.


\end{abstract}

\section{Introduction}

Since the Searle \& Sargent (1972) paper identifying blue compact galaxies
(BCGs), that is, low-mass galaxies showing  emission line spectra
characteristic of H{\sc ii} regions, intense star formation (SF), and oxygen
abundances of 1/50 -- 1/3 solar\footnote{12+log(O/H)$_{\odot}$ = 8.92
(Anders \& Grevesse \cite{Anders89}).}, such objects
have been considered as young  galaxies undergoing
one of their first star formation bursts.
I~Zw~18, a BCG with the lowest known oxygen abundance among the
galaxies (O/H $\sim$ 1/50 (O/H)$_{\odot}$, Searle \& Sargent \cite{Searle72};
Izotov \& Thuan \cite{IT99}),  has been suggested as a candidate to be a
truly-local young galaxy, experiencing
its first short SF episode. 
The second candidate young galaxy, SBS~0335--052E, with an oxygen abundance of
1/41 (O/H)$_{\odot}$ (Melnick et al. \cite{Melnick92}; Izotov et al.
\cite{Izotov97a}; Lipovetsky et al. \cite{Lipovetsky99}) was discovered
18 years later by Izotov et al. (\cite{Izotov90}).
With only two probable examples, we must be extremely lucky to be witnessing
local galaxy formation. The proximity of these probable
young galaxies allows one to study their properties in detail
and to set important constraints on models of galaxy formation.
Such studies are important for understanding the nature 
of very faint and compact probable primeval galaxies at high redshifts.
Most of such galaxies at $z = 3-5$ were discovered only recently
(e.g. Steidel et al. \cite{Steidel96}; Dey et al. \cite{Dey98}),
and  it seems that the majority  of them are already rather evolved systems.
Moreover, the local candidate young galaxies are at least one order of
magnitude less massive than the faintest candidate young galaxies
at high redshifts, and represent the range of baryon mass
(10$^{8}$--10$^{9}$ $M_{\odot}$) within which possibly most of primeval
galaxies have formed (e.g. Rees \cite{Rees88}).

Evidence for the existence of old low-mass stellar populations was obtained 
in the last 25 years for most of the studied BCGs (Thuan \cite{Thuan83};
Loose \& Thuan~\cite{Loose86}). Moreover, no conclusive answer
has been reached yet about the youth of the few most metal-poor BCG.
However, some observational data have been
collected lately, which apparently support young ages for these BCGs.
Among them we point out:

a) Extremely low abundances of heavy elements in H{\sc ii} regions
surrounding  young
clusters, consistent with theoretical expectations of ``metal'' yield 
during a first SF event ($Z$ $<$ 1/20 $Z_{\odot}$) (e.g., Pilyugin 
\cite{Pil93});

b) Very blue colours outside the location of the current SF burst, consistent
with a lack of stars older than 100 Myr (Hunter \& Thronson \cite{Hunter95};
Papaderos et al. \cite{Papa98}). While the recent analysis of HST data for
I~Zw~18 by Aloisi et al. (\cite{Aloisi99}) suggests an age of 1 Gyr
for the underlying stellar population of the galaxy,   Izotov et al.
(\cite{Izotov2000}) argue that a self-consistent treatment of all data favours
a significantly larger distance to I~Zw~18 then adopted by Aloisi et al.,
and a 100 Myr stellar population;

c) A large amount of neutral gas, making  up 99\% of all baryonic
(luminous) mass (van Zee et al. \cite{vanZee98}; Pustilnik et al.
\cite{Pus2000});

d) Practically zero metallicity for this H{\sc i} gas,
e.g., (O/H) $<$ 3$\times$10$^{-5}$(O/H)$_{\odot}$, as reported for 
SBS~0335--052E (Thuan \& Izotov \cite{TI97}).
This emphasizes either an extremely slow evolution on these systems,
or a very recent onset of metal production.
The latter suggests that the neutral gas clouds in these galaxies are composed
of pregalactic material not yet polluted by stellar nucleosynthesis products.

It was suggested recently by Izotov \& Thuan (\cite{IT99}), 
from the analysis of carbon and nitrogen abundances, that several 
BCGs with O/H $<$ 1/20 (O/H)$_{\odot}$ in H{\sc ii} regions
are good candidate galaxies with a recent first 
SF episode. Until now, less than ten such galaxies with good abundance
determinations are  known. Even though the point on the existence of 
truly young
local galaxies is debatable (see, e.g., Kunth \& \"Ostlin~\cite{Kunth99}),
the importance of studies of extremely metal-poor galaxies is undoubtful, 
since
they best approximate the properties of primeval galaxies at
large redshifts.

In this paper we describe the data obtained for the third most 
metal-deficient galaxy, HS~0822+3542 with O/H = 1/36 (O/H)$_{\odot}$.
This is one of the {\it nearest}, and at the same time the
{\it dimmest} candidate young galaxy known. 

\begin{figure}
    \psfig{figure=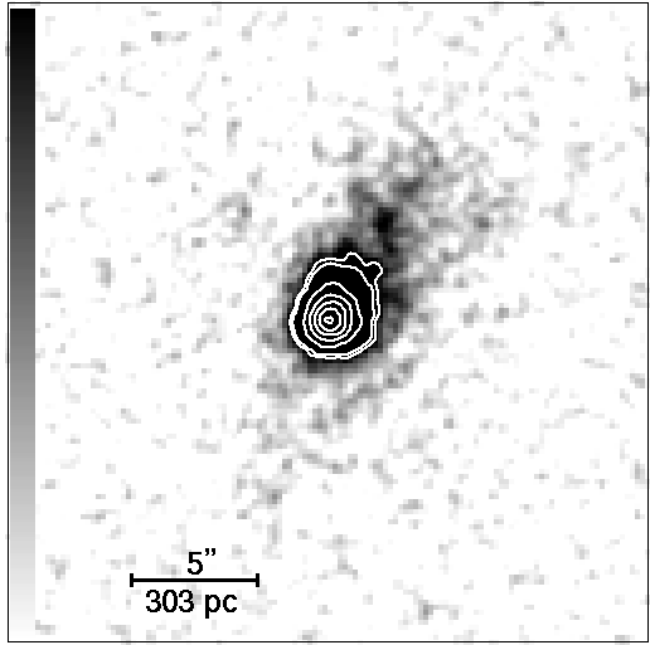,angle=0,width=8.5cm,bbllx=70pt,bblly=300pt,bburx=520pt,bbury=750pt}
    \caption{$R$-band image of
    HS~0822+3542. North is to the top, east is to the left. Low-contrast
    filaments   NW of the main bright body indicate gas
    structures typical of high-velocity ejecta.
    At the adopted distance of 12.5 Mpc, 1$^{\prime\prime}$ = 60.6 pc. 
        Surface brightness
    isophotes (white contours) are superposed on the central bright region
    to show the morphology of supergiant H {\sc ii} region.}
     \label{fig:HS_direct}
\end{figure}

\begin{table}[h]
\centering{
\caption{Main parameters of the young galaxy candidates}
\label{tab:Main_par}
%\footnotesize{
\begin{tabular}{lrrr} \hline\hline
%\multicolumn{1}{c}{} & \multicolumn{3}{c}{}   \\
Parameter                     & 0822+3542$^{1}$   & IZw18                 & 0335--052E    \\ \hline

$B_{\rm tot}$                     &17.92$\pm$0.07~~~  & 16.21$^{2}$  & 17.00$\pm$0.02$^{3}$   \\
$(B-V)_{\rm tot}$                 &0.32$\pm$0.08~~~   & 0.18$^{2}$   &  0.31$\pm$0.05$^{3}$   \\
$(V-R)_{\rm tot}$                 &0.17$\pm$0.09~~~   & 0.41$^{4}$    &  0.15$\pm$0.05$^{3}$   \\
$V_{\rm Hel}$ (km s$^{-1}$)       &  732$\pm$6~~~     & 751$\pm$2$^{5}$   & 4043$\pm$5$^{6}$  \\
$D_{\rm Vir}$(Mpc)                & 12.5~~~           & 14.2$^{1}$    & 52.8$^{6}$   \\
     $E(B-V)$                     &     0.047$^{11}$  &     0.032$^{11}$  &     0.047$^{11}$     \\
$M_{B}$$^{\dag}$              &  --12.7~~~        & --14.64~~      & --16.8$^{3}$ \\
Angular size (\arcsec)$^{\ddag}$ & 14.8$\times$7.4~~~  & 22$\times$15$^{1}$    & 14$\times$10$^{3}$  \\
Optical size (kpc)               & 0.90$\times$0.45~~~& 1.5$\times$1.0~        & 3.7$\times$2.6$^{3}$ \\
12+log(O/H)  \                & 7.35~~~           & 7.16$^{7}$  & 7.29$^{8}$~ \\
$T_{\rm e}$(O {\sc iii}) (K)\              & 20,350~~~         & 19,600$^{7}$~& 19,300$^{8}$~ \\
H {\sc i} flux$^{*}$                & 0.68$\pm$0.07~~~  & 2.97$^{9}$  & 2.46$^{6}$  \\
$W_{50}$ (km s$^{-1}$)        & 42$\pm$5~~~       & 49$^{9}$    & 83$^{5}$    \\
$W_{20}$ (km s$^{-1}$)        & 58$\pm$8~~~       & 84$^{9}$    & 105$^{5}$   \\
$M$(H{\sc i}) (10$^{8}$$M_{\odot}$)   & 0.24~~~           & 1.41$^{9}$  & 16.2$^{6}$ \\
$M$(H{\sc i})/$L_{B}$$^{**}$         & 1.40~~~           & 1.40~     & 2.3~~       \\
SFR ($M_{\odot}$/year)         & 0.007~~~          & 0.04$^{10}$  & 0.4$^{10}$    \\
\hline\hline
\multicolumn{4}{l}{$B_{\rm tot}$ -- total blue magnitude; $M_{B}$ -- absolute blue mag.} \\
\multicolumn{4}{l}{$L_{B}$ -- total blue luminosity. $^{*}$\ Units of Jy km s$^{-1}$;}\\
\multicolumn{4}{l}{$^{**}$ In units of ($M$/$L_{B}$)$_{\odot}$; $^{\dag}$ With the Galaxy extinction}  \\
\multicolumn{4}{l}{~$A_{B}$ = 0.20, 0.14, 0.20, respectively, } \\
\multicolumn{4}{l}{{corresponding to $E(B-V)$ in the previous line};} \\
\multicolumn{4}{l}{$^{\ddag}$\ $a \times b$ at surface brightness 
$\mu_{B}$=25 mag arcsec$^{-2}$.} \\
\multicolumn{4}{l}{{\bf References}: $^1$This paper; $^{2}$Mazzarella \& Boroson (\cite{Maz93});} \\
\multicolumn{4}{l}{ $^{3}$Papaderos et al. (\cite{Papa98}); $^{4}$Huchra (\cite{Huchra77}); $^{5}$Thuan et al.}\\
\multicolumn{4}{l}{(\cite{TLMP99}); $^{6}$Pustilnik et al. (\cite{Pus2000}); $^{7}$Izotov \& Thuan (\cite{IT98});} \\
\multicolumn{4}{l}{$^{8}$Izotov \& Thuan (\cite{IT99}); $^{9}$van Zee et al. (\cite{vanZee98});}  \\
\multicolumn{4}{l}{$^{10}$ Thuan et al. (\cite{TIL97}); $^{11}$ Schlegel et al. (\cite{Schlegel98}) } \\
\end{tabular}
}
%}
\end{table}

\section{Observations and data reduction}

A new extremely metal-poor BCG, HS~0822+3542, was discovered on April 5, 1998,
during observations with the 6\,m telescope of the Special Astrophysical
Observatory (SAO) of the Russian Academy of Sciences (Pustilnik et al.
\cite{Pus99b}),
in the framework of the Hamburg/SAO survey for emission-line galaxies
(Ugryumov et al. \cite{Ugryumov99}). Its J2000 coordinates are: R.A. =
08$^{\rm h}$25$^{\rm m}$55\fs0, Dec. = +35$^{\circ}$32\arcmin31\arcsec.
The main parameters of the galaxy are presented
in Table~\ref{tab:Main_par}.
Here we present new optical spectroscopic and photometric, and H{\sc i} 21 cm
radio observations to study the properties of this galaxy.

\begin{figure*}[hbtp]
\begin{center}
    \psfig{figure=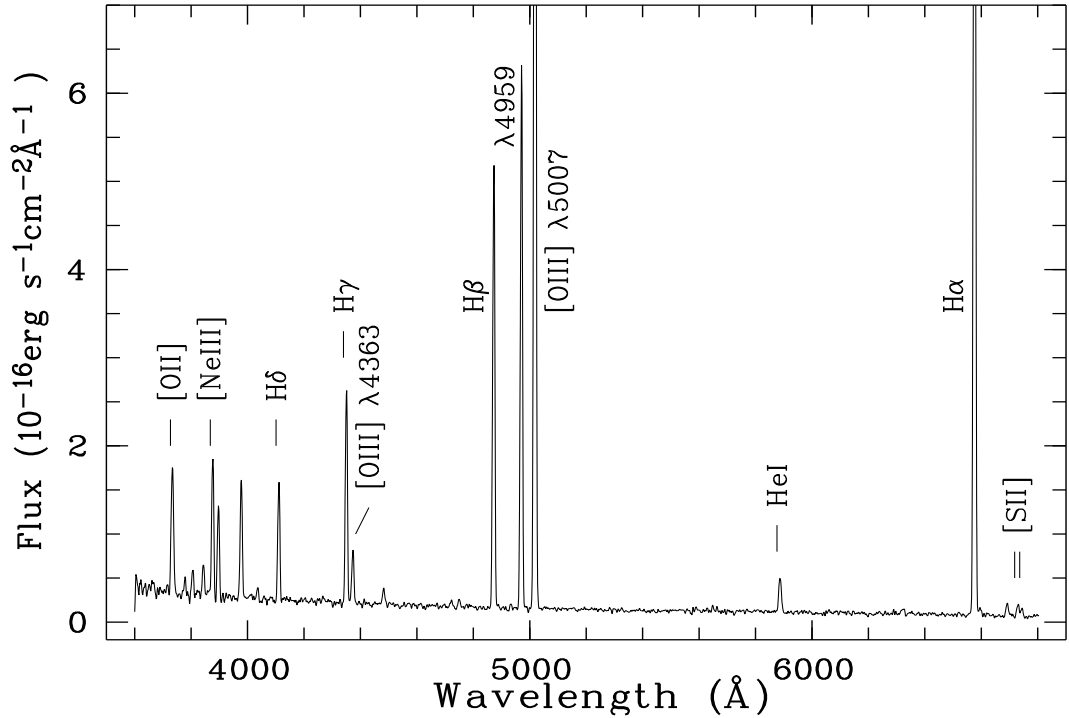,angle=270,width=16cm,height=10cm}
    \caption{The observed NOT spectrum of
     HS~0822+3542 in the aperture
     1\farcs3$\times$0\farcs8 with nebular emission lines identification.}
    \label{fig:HS_spec}
\end{center}
\end{figure*}

\subsection{Long-slit spectroscopy}

\subsubsection{Observations}

Optical spectra of HS~0822+3542 were obtained 
with the 6\,m telescope on April 6,
1998 with  the spectrograph SP-124 equipped with a Photometrics
CCD-detector ($24\times24\mu$m pixel size) and operating at
the Nasmyth-1 focus.
The grating B0 with 300 grooves/mm provides dispersion of 4.6~\AA\ pixel$^{-1}$
and spectral resolution of about 12~\AA\ at first order.
A long slit with a size 2\arcsec$\times$40\arcsec\ was used.
The scale along the slit was 0\farcs4 pixel$^{-1}$.
The total spectral range covered was $\lambda\lambda$\,3700--8000~\AA.

High S/N ratio long-slit
spectroscopy was conducted with the 2.5\,m Nordic Optical Telescope (NOT)
on May 27 and 28, 1998.
We used the spectrograph ALFOSC equipped with a
Loral (W11-3AC) CCD, with a
1\farcs3$\times$400\arcsec\ slit, and 
grisms \#6 and \#7 (110~\AA/mm), which provide a spectral dispersion
of 1.5~\AA\ pixel$^{-1}$
and a resolution of about 8~\AA\ (FWHM).
The spectral range was $\lambda\lambda$\,3200--5500~\AA\ for grism \#6 and
$\lambda\lambda$\,3800--6800~\AA\ for grism \#7. The spatial resolution was of
0\farcs189 per pixel.
The total exposure time for grism \#6 was 60 min, split into three
20 min exposures, and 40 min for grism \#7, split into two 
20 min exposures.
The slit, centred on the brightest knot
(see in Fig.~\ref{fig:HS_direct} the $R$-band image of the
galaxy), was oriented in the N-S direction.
Spectra of He-Ne comparison lamp
were obtained after each exposure for wavelength calibration, and three
spectrophotometric standard stars Feige~34, HZ~44 and 
\mbox{BD+33$^{\circ}$2642}, 
were observed for flux calibration. 
The seeing during the spectral observations was $\approx$ 0\farcs8.

\subsubsection{Reduction of long-slit spectra}

Data reduction was carried out using the
MIDAS\footnote{MIDAS is an acronym for the European Southern Observatory
package --- Munich Image Data Analysis System.}
software package (Grosb{\o}l \cite{Grosbol89}).
Procedures included bias subtraction,
cosmic-ray removal and flat-field correction.
The flat-field correction was produced with the normalization algorithm
described by Shergin et al. (\cite{SKL96}).
After the wavelength mapping and night sky subtraction,
each 2D frame was corrected for atmospheric extinction and was flux
calibrated.
To derive the sensitivity curves, we used the
spectral energy distributions of the standard stars from
Bohlin (\cite{Bohlin96}). Average sensitivity curves for grisms were produced
for each observing night.
R.m.s. deviations between the average and individual sensitivity
curves are $\approx$1.5\%, with the maximum deviations of $\approx$4\%
in the spectral region 3800$\div$4000 \AA.

The 2D flux-calibrated spectra were then corrected for atmospheric
dispersion (see Kniazev et al.\, \cite{Kniazev2000}) and averaged.
Finally, the 1D averaged spectrum was extracted from a 
0\farcs8 region along the slit, where
\mbox{$I$($\lambda$4363~\AA)} $>$ 2$\sigma$ ($\sigma$ is the dispersion of
a noise statistics around this line) (see Fig.~\ref{fig:HS_spec}).

Redshift and line fluxes
were measured applying Gaussian fitting.
For H$\alpha$,
[N{\sc ii}]\,$\lambda\lambda$\,6548,6583~\AA\ and
[S{\sc ii}]\,$\lambda\lambda$\,6716,6731~\AA\ a deblending procedure
was used, assuming gaussian profiles with the same FWHM  as for single 
lines.  
The errors of the line intensities in Table~\ref{t:Intens}
take into account
the Poisson noise statistics and the noise statistics in the continuum
near each line, and include uncertainties of data reduction.
These errors have been propagated to calculate element abundances.

The observed emission line intensities $F(\lambda)$, and those corrected
for interstellar extinction and underlying stellar absorption $I(\lambda)$
are presented in Table~\ref{t:Intens}. All lines have been normalized 
to the H$\beta$ intensity. The H$\beta$ equivalent width $EW$(H$\beta$),
the absorption equivalent widths $EW$(abs) of the Balmer lines,
the H$\beta$ flux, and the
extinction coefficient $C$(H$\beta$) (this is a sum of internal
extinction in HS~0822+3542 and foreground one in the Milky Way)
are also shown there.

For the simultaneous derivation  of $C$(H$\beta$) and $EW$(abs),  and
to correct for extinction, we used a procedure described in detail
in Izotov, Thuan \& Lipovetsky (\cite{Izotov94}).
The abundances of the ionized species and the total abundances 
of  O, Ne, N, S, and He  have been obtained following
Izotov, Thuan \& Lipovetsky (\cite{Izotov94}, \cite{Izotov97b})
and Izotov \& Thuan (\cite{IT99}).
 
\subsection{CCD photometry}

\subsubsection{Observations}

CCD images in Bessel $BVR$ filters were obtained with the NOT
and ALFOSC on 1998, May 28.
The same 2k$\times$2k Loral (W11-3AC) CCD was used, with
a plate scale of 0\farcs189$\times$0\farcs189 and a 
6\farcm5$\times$6\farcm5  field of view.
Exposures of  900 s in $B$ and 600 s in both
$V$ and $R$ were obtained under photometric conditions
but no photometric calibration was performed.
The seeing FWHM  was  1\farcs25.
Dr. A. Kopylov (SAO RAS) kindly obtained short CCD images in $BVR$ with
the 1\,m telescope of SAO.
These observations were used to calibrate  the NOT data.
Photometric calibration was provided by observations of the
stars \#4, \#7 and \#10 from the field of OJ~287 (Firucci \& Tosti
\cite{Firucci96}; Neizvestny \cite{Neizvestny95}).

\subsubsection{Reduction of photometric data}

All primary data reduction was done with MIDAS.
The frames were corrected  for bias,
dark, and flat field in the same way as for reduction of the 2D frames of NOT
long-slit spectra.
Aperture photometry was performed on the standard star frames using the
{\tt MAGNITUDE/CIRCLE} task, with the same aperture for all stars.
The instrumental magnitudes were transformed to the standard
photometric system magnitudes via secondary local standards, calibrated
with the 1\,m telescope of SAO.
The final zero-point uncertainties of the transformation were
$\sigma_{B}$ = $0\fm06$ in $B$, $\sigma_{V}$ = $0\fm05$ in $V$,
and $\sigma_{R}$ = $0\fm08$ in $R$.

For the creation of the sky background, we used
the dedicated software for adaptive filtering developed at the
Astrophysical Institute of Potsdam (Lorenz et al. \cite{Lor93}).

The photometry of extended objects was carried out with
the IRAF\footnote{IRAF: the Image Reduction and Analysis Facility is
distributed by the National Optical Astronomy Observatories, which is
operated by the Association of Universities for Research in Astronomy,
In. (AURA) under cooperative argeement with the National Science
Foundation (NSF).} software package.
Elliptical fitting was performed with the {\tt ELLIPSE} task
in the {\tt STSDAS} package.
To construct a surface brightness (SB) profile  we used the equivalent
radius as the geometrical average, $R^*$=$\sqrt{ab}$.
The SB profiles were decomposed into two components:
one with a  gaussian distribution
in the  central part and the second being an exponential disc.
The final function has the form:
\begin{equation}
 I = I_{\rm E,0} \exp\left(-\frac{R^*}{\alpha_{\rm E}}\right) + 
I_{\rm G,0} 
\exp\left[-\ln 2  \left(2 \frac{R^*}{\alpha_{\rm G}}\right)^2~\right]
\end{equation}

\begin{figure}[hbtp]
    \hspace*{-0.4cm}\psfig{figure=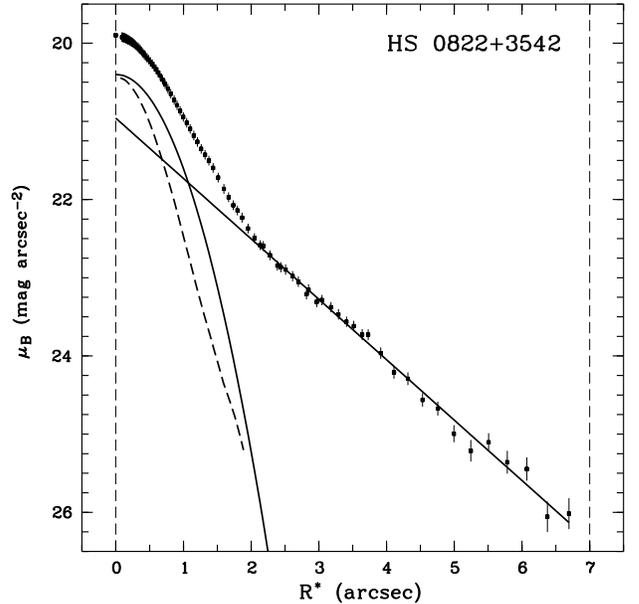,angle=270,width=12.5cm}
    \caption{The surface brightness profile  of HS~0822+3542 in
         $B$-band. Error bars correspond to 2$\sigma$ uncertainties.
         By the solid lines the decomposition of SB profile is shown
         to the underlying exponential disc and the central gaussian
         component corresponding to the SF burst region. The decomposition
         was performed with the weights.
         Thick dashed line shows
         the PSF derived on nearby stars, with FWHM = 1\farcs25.
     Two vertical dashed lines mark the region over which
         the decomposition was performed.
     }
    \label{fig:SBP}
\end{figure}

For the profile decomposition, the {\tt NFIT1D} task of the {\tt STSDAS}
was used with weights inversely proportional to the  accuracy of
the surface brightness profile.  The final photometric errors take into
account the instrumental errors and the error of transformation
to the standard photometrical system. To check correctness of disc parameters
we performed separately the fitting of external part of SB profile, corresponding
to the region in Fig.~\ref{fig:SBP} from $R^*$ of 3\arcsec\ to 6\arcsec.
The derived disc parameters were the same within the cited errors.

For consistency with previous works (e.g. Telles et al. \cite{Telles97},
Papaderos et al. \cite{Papa96}, \cite{Papa98}, Doublier et al. \cite{Doublier99a})
the parameters of the disc have been obtained by fitting again the SB versus
the equivalent radius, but we note that 
 the obtained values should be taken with caution, because of
ellipticity variations.

\begin{figure}[hbtp]
    \hspace*{-0.3cm}\psfig{figure=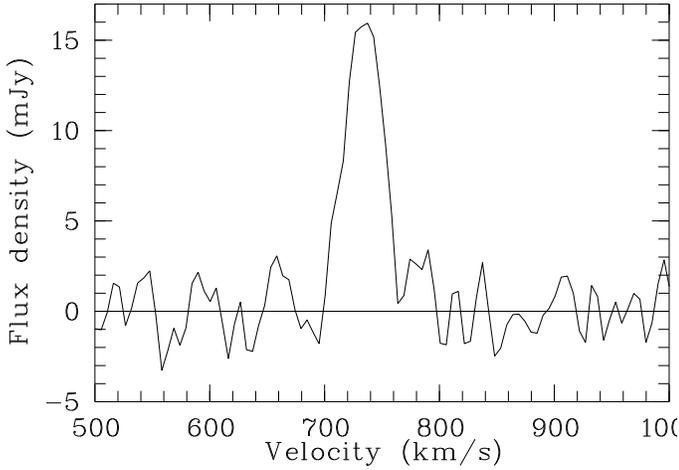,angle=270,width=9.5cm}
    \caption{H{\sc i} 21 cm profile of HS~0822+3542.
     }
    \label{fig:HIprof}
\end{figure}

\subsection{ {\rm H}{\sc i} 21 cm observations and data reduction}

H{\sc i} line observations were carried out in July  1998 and
in February 1999 with the Nan\c {c}ay\footnote{The Nan\c {c}ay 
Radioastronomy Station is part of the Paris Observatory and is operated 
by the Minist\`ere de l'Education Nationale and Institut des Sciences 
de l'Univers of the Centre National de la Recherche Scientifique.} 
300m radio telescope (NRT). The NRT  has a half-power beam width of
3\farcm7 (EW) $\times$ 22\arcmin\ (NS) at the declination
Dec. = 0$^\circ$.

Since HS~0822+3542 had a known optical redshift, 
we split the 1024-channel autocorrelator in two halves and used
a dual-polarization receiver  to increase the S/N ratio.
Each correlator segment covered a 6.4 MHz bandwidth, corresponding to 
a 1350 km s$^{-1}$ velocity coverage, and was centred at the frequency
corresponding to the optical redshift. 
The channel spacing was 2.6 km s$^{-1}$ before smoothing and the effective
resolution after averaging pairs of adjacent channels and Hanning smoothing
was 10.6 km s$^{-1}$. The system temperature of the receiver was 
$\approx$ 40~K in the horizontal and vertical  linear polarizations.
The gain of the telescope was 1.1~K/Jy at the declination 
Dec. = 0$^\circ$. The observations were made in the standard total
power (position switching) mode with 1-minute on-source
and 1-minute off-source integrations.

The data were reduced using the NRT standard programs DAC and SIR, 
written by the telescope's staff. Both H and V polarization spectra were
calibrated and processed independently, and were finally averaged
together. Error estimates were calculated following Schneider et al.
(\cite{Schneider86}).
With an integration time of 210 minutes, the r.m.s. noise is of 1.4 mJy after
smoothing. HS~0822+3542 is detected with S/N=11. The spectrum is
presented in Fig.~\ref{fig:HIprof}.

\begin{table}[h]
     \centering{
\caption{Line intensities in HS~0822+3542}
\label{t:Intens}
\begin{tabular}{lcc} \hline \hline
$\lambda_{0}$(\AA) Ion                  &$F$($\lambda$)/$F$(H$\beta$)
&$I$($\lambda$)/$I$(H$\beta$)   \\ \hline
3727\ [O {\sc ii}]              & 0.331 $\pm$0.013 &  0.331 $\pm$0.014   \\
3835\ H9                        & 0.077 $\pm$0.005 &  0.080 $\pm$0.007   \\
3868\ [Ne {\sc iii}]            & 0.317 $\pm$0.018 &  0.318 $\pm$0.018   \\
3889\ H8\ +\ He {\sc i}         & 0.199 $\pm$0.009 &  0.203 $\pm$0.010   \\
4026\ He {\sc i}                & 0.026 $\pm$0.005 &  0.026 $\pm$0.005   \\
4101\ H$\delta$                 & 0.271 $\pm$0.011 &  0.274 $\pm$0.012   \\
4340\ H$\gamma$                 & 0.479 $\pm$0.018 &  0.481 $\pm$0.018   \\
4363\ [O {\sc iii}]             & 0.123 $\pm$0.007 &  0.123 $\pm$0.008   \\
4471\ He {\sc i}                & 0.039 $\pm$0.005 &  0.039 $\pm$0.005    \\
4861\ H$\beta$                  & 1.000 $\pm$0.033 &  1.000 $\pm$0.034   \\
4922\ He {\sc i}                & 0.009 $\pm$0.003 &  0.009 $\pm$0.003   \\
4959\ [O {\sc iii}]             & 1.192 $\pm$0.042 &  1.190 $\pm$0.042   \\
5007\ [O {\sc iii}]             & 3.550 $\pm$0.121 &  3.542 $\pm$0.121   \\
5876\ He {\sc i}                & 0.098 $\pm$0.005 &  0.097 $\pm$0.005   \\
6300\ [O {\sc i}]               & 0.004 $\pm$0.003 &  0.004 $\pm$0.003   \\
6312\ [S {\sc iii}]             & 0.010 $\pm$0.003 &  0.010 $\pm$0.003   \\
6548\ [N {\sc iii}]             & 0.005 $\pm$0.004 &  0.005 $\pm$0.004   \\
6563\ H$\alpha$                 & 2.743 $\pm$0.087 &  2.729 $\pm$0.094   \\
6584\ [N {\sc ii}]              & 0.015 $\pm$0.012 &  0.015 $\pm$0.012   \\
6678\ He {\sc i}                & 0.033 $\pm$0.004 &  0.033 $\pm$0.004   \\
6717\ [S {\sc ii}]              & 0.029 $\pm$0.004 &  0.028 $\pm$0.004   \\
6731\ [S {\sc ii}]              & 0.018 $\pm$0.004 &  0.018 $\pm$0.004   \\
                     & & \\
$C$(H$\beta$)\ dex                        &\MC {2}{c}{0.005$\pm$0.04}                   \\
$F$(H$\beta$)                             &\MC {2}{c}{0.47$\times10^{-14}$\ erg\ s$^{-1}$cm$^{-2}$}        \\
$EW$(H$\beta$)~\AA                        &\MC {2}{c}{292$\pm$3}                       \\
$EW$(abs)~\AA                             &\MC {2}{c}{0.6$\pm$0.7}                      \\
\hline\hline
\end{tabular}
}
\end{table}

\begin{table*}[hbtp]
    \centering{
\caption{Abundances in HS 0822+3542, SBS 0335--052 and I Zw 18}
\label{t:Chem}
\begin{tabular}{lccccc} \hline \hline
Value                                & HS~0822+3542     & SBS~0335--052E$^{1,2}$ &SBS~0335--052W$^3$&I~Zw~18NW$^4$&I~Zw~18SE$^4$ \\ \hline
$T_{\rm e}$(O {\sc iii})(K)\                      & 20,360$\pm$850~~ & 20,300$\pm$300~~& 17,200$\pm$500 & 19,700$\pm$200~~& 18,800$\pm$400~~ \\
$T_{\rm e}$(O {\sc ii})(K)\                       & 15,790$\pm$600~~ & 15,800$\pm$200~~& 14,700$\pm$400 & 15,600$\pm$150~~& 15,300$\pm$300~~\\
$T_{\rm e}$(S {\sc iii})(K)\                      & 18,600$\pm$700~~ & 18,500$\pm$200~~& 16,000$\pm$400 & 18,000$\pm$200~~& 17,300$\pm$350~~\\
$N_{\rm e}$(S {\sc ii})(cm$^{-3}$)\               & $<$10$\pm^{40}_{10}$ & 524$\pm$204& 10& 90& 10   \\
O$^{+}$/H$^{+}$($\times$10$^5$)\     & 0.247$\pm$0.025  & 0.20$\pm$0.1& 0.60$\pm$0.05& 0.22$\pm$0.01& 0.49$\pm$0.03   \\
O$^{++}$/H$^{+}$($\times$10$^5$)\    & 1.967$\pm$0.190  & 1.70$\pm$0.1& 1.08$\pm$0.08& 1.16$\pm$0.03& 1.04$\pm$0.06    \\
O/H($\times$10$^5$)\                 & 2.214$\pm$0.191  & 1.90$\pm$0.1& 1.68$\pm$0.10& 1.45$\pm$0.03& 1.54$\pm$0.07     \\
12+log(O/H)\                         & ~7.35$\pm$0.04   &  ~7.29$\pm$0.01& ~7.22$\pm$0.03& 7.16$\pm$0.01& 7.19$\pm$0.02  \\
N$^{+}$/H$^{+}$($\times$10$^7$)\     & 0.992$\pm$0.770  & 0.60$\pm$0.01& 1.72$\pm$0.15& 0.64$\pm$0.02& 1.43$\pm$0.08      \\
ICF(N)\                              & 8.962            & 8.66 & 2.81& 6.59& 3.14                 \\
log(N/O)\                            &--1.40$\pm$0.34   &--1.58$\pm$0.03&--1.54$\pm$0.06&--1.56$\pm$0.02&--1.53$\pm$0.04   \\
Ne$^{++}$/H$^{+}$($\times$10$^5$)\   & 0.355$\pm$0.041  & 0.27$\pm$0.05& 0.24$\pm$0.03& 0.19$\pm$0.01& 0.23$\pm$0.03        \\
ICF(Ne)\                             & 1.126            & 1.13 & 1.55& 1.25& 1.48                 \\
log(Ne/O)\                           &--0.74$\pm$0.07   &--0.80$\pm$0.03&--0.65$\pm$0.06&--0.80$\pm$0.01&--0.65$\pm$0.04   \\
S$^{+}$/H$^{+}$($\times$10$^7$)\     & 0.424$\pm$0.057  & 0.40$\pm$0.1 &1.09$\pm$0.11& 0.35$\pm$0.01& 0.67$\pm$0.03         \\
S$^{++}$/H$^{+}$($\times$10$^7$)\    & 2.815$\pm$0.873  & 1.90$\pm$0.3 &2.15$\pm$0.64& 1.96$\pm$0.21& 2.07$\pm$0.30          \\
ICF(S)\                              & 2.26             & 2.21&1.29& 1.82& 1.32                  \\
log(S/O)\                            &--1.48$\pm$0.09   &--1.59$\pm$0.04&--1.60$\pm$0.08&--1.55$\pm$0.03&--1.63$\pm$0.04   \\
$Y$(mean)\                             & ~0.255$\pm$0.013 & ~0.249$\pm$0.006& 0.238$\pm$0.005& 0.217$\pm$0.005& 0.242$\pm$0.009  \\
\hline\hline
\multicolumn{6}{l}{
{\bf References}: $^{1}$ Izotov et al. (\cite{Izotov97a});
$^{2}$ Izotov \& Thuan (\cite{IT99}); $^{3}$ Lipovetsky et al. 
(\cite{Lipovetsky99}); $^{4}$ Izotov \& Thuan (\cite{IT98}).} \\
\end{tabular}
}
\end{table*}

\begin{table*}[hbtp]
    \centering{
\caption{Structural properties of HS~0822+3542 in $B$,$V$,$R$-bands}
\label{tab:struct_par}
\begin{tabular}{ccccccccc} \hline \hline
Band & $\mu_{\rm E,0}$      &$\alpha_{\rm E}$    & $\mu_{\rm G,0}$      & $\alpha_{\rm G}$  & $P_{25}$ & $E_{25}$&   $m_{\rm LSB}\infty$&   $m_{\rm SF}\infty$  \\
     & mag arcsec$^{-2}$&  arcsec      & mag arcsec$^{-2}$&    arcsec     &   pc     &   pc    &     mag          &     mag           \\
     &    (1)           &  (2)         &     (3)          &    (4)        &  (5)     &  (6)    &      (7)         &      (8)          \\ \hline
$B$  & 20.96$\pm$0.04   &1.41$\pm$0.02 & 20.40$\pm$0.02   & 1.58$\pm$0.02 &  118     &  318    &     18.22        &     19.46         \\
$V$  & 20.86$\pm$0.04   &1.42$\pm$0.03 & 19.92$\pm$0.02   & 1.61$\pm$0.03 &  126     &  333    &     18.10        &     18.68         \\
$R$  & 20.57$\pm$0.08   &1.38$\pm$0.04 & 19.79$\pm$0.03   & 1.61$\pm$0.03 &  129     &  342    &     17.88        &     18.60         \\
\hline\hline
\multicolumn{9}{p{14cm}}{(1) Central surface brightness of the LSB component                                                                                                                                                                                                                                                                      obtained from the decomposition of each SB profile, weighted
          by its photometric  uncertainties. }                                            \\
\multicolumn{9}{p{14cm}}{(2) Exponential scale length of the LSB component.} \\
\multicolumn{9}{p{14cm}}{(3) Central surface brightness of the gaussian (SF burst) component.
        }\\
\multicolumn{9}{p{14cm}}{(4) Effective size (FWHM) of the gaussian (SF burst) component.}              \\
\multicolumn{9}{p{14cm}}{(5) Linear extent of the luminous component in
             excess of the LSB component at a surface
             brightness level of 25 mag arcsec$^{-2}$.}\\
\multicolumn{9}{p{14cm}}{(6) Linear extent of the LSB component at a surface
             brightness level of 25 mag arcsec$^{-2}$.}\\
\multicolumn{9}{p{14cm}}{(7) Total apparent magnitude of the LSB component
             estimated by extrapolation of the exponential
             fitting law to $R^*$ = $\infty$ (equation 2).}\\
\multicolumn{9}{p{14cm}}{(8) Total apparent magnitude of the SF component.}\\
\end{tabular}
}
\end{table*}

\begin{figure}[hbtp]
    \hspace*{-0.4cm}\psfig{figure=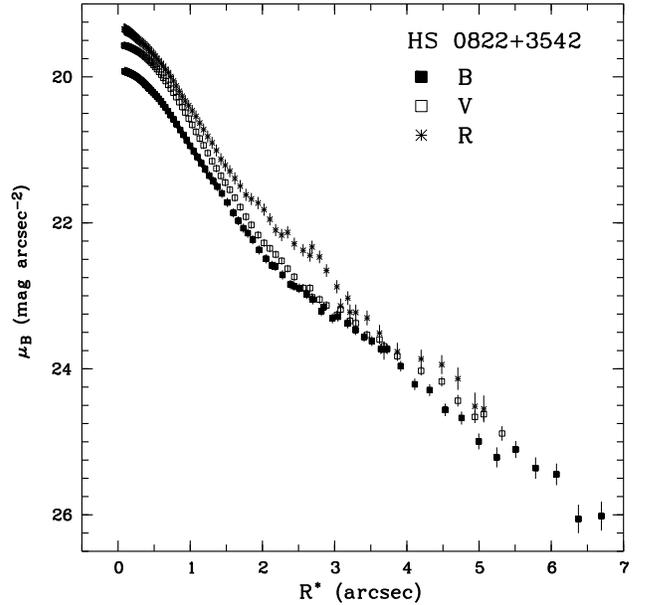,angle=270,width=12.5cm}
    \caption{The surface brightness profiles  of HS~0822+3542 in
         $B$, $V$ and $R$-bands.
         Error bars correspond to 2$\sigma$ uncertainties.
     }
    \label{fig:SBPs}
\end{figure}

\begin{figure}[hbtp]
    \hspace*{-0.4cm}\psfig{figure=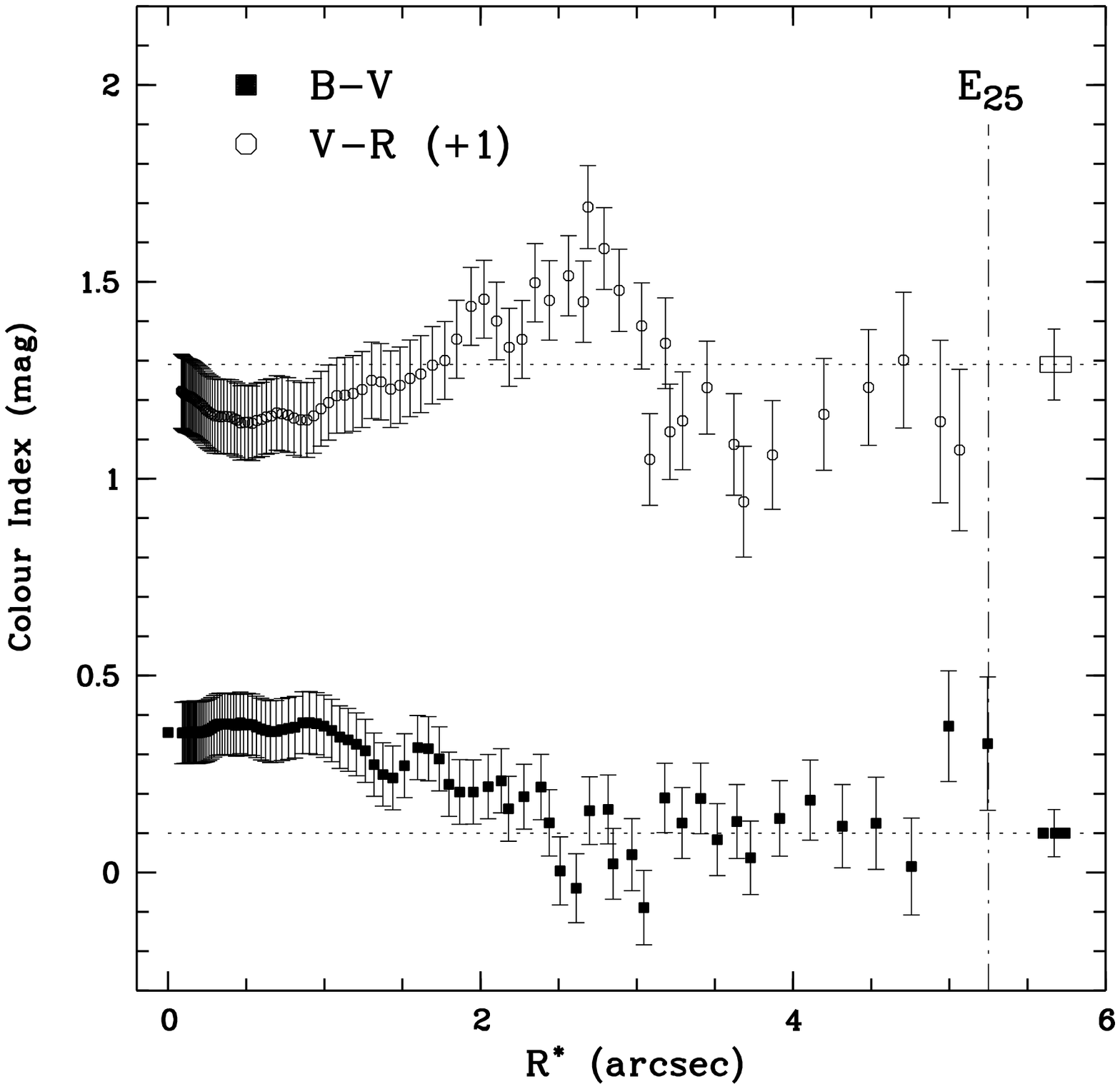,angle=270,width=12.5cm}
    \caption{($B-V$) and ($V-R$) radial colour profiles of HS~0822+3542.
      Radially averaged ($B-V$) and ($V-R$) colour profiles computed
     by subtraction of surface brightness profiles displayed in 
     Fig.~\ref{fig:SBPs}.
     The mean colours of the exponential disc are shown by the filled and
     open rectangulars with error bars.
     The isophotal radius $E_{25}$ of the LSB host at 25 $B$ mag arcsec$^{-2}$
     is indicated.
     }
    \label{fig:phot_colours}
\end{figure}

\section{Results and discussion}

The main parameters of HS~0822+3542, along with those of the two earlier known
local candidate young galaxies, are presented in Table~\ref{tab:Main_par}.

The distances to all three galaxies for the sake of homogeneity are
derived from the radial velocities.
For the distance-dependent parameters we used the
distances $D_{\rm Vir}$ derived from the quoted heliocentric H{\sc i} velocities
$V_{\rm HI}$, accounting for the Galaxy motion relative to the centroid
of the Local Group of 220~km s$^{-1}$, for Virgocentric infall
(Kraan-Korteweg \cite{Kraan86}), and assuming a
Hubble constant $H_{0}$ = 75~km s$^{-1}$Mpc$^{-1}$.
Note that HS~0822+3542 is the
{\it nearest} galaxy among the extremely metal-deficient galaxies 
shown in Table~\ref{tab:Main_par}. 
However, such distance estimates for two galaxies, HS~0822+3542 and I~Zw~18
with small radial velocities are rather uncertain. In particular, for
I Zw 18 different distance determinations from the literature 
based on the different observational data, range from 10.9 Mpc to 20 Mpc. 
To resolve this disagreement the deep HST imaging
of I~Zw~18 is vital in order to detect the tip of the red giant
branch  stars and to measure the distance directly.
Unfortunately, such observations have not yet been done.
The same relates to HS~0822+3542.
Keeping in mind this
possible source of uncertainty which may change integrated characteristics
of HS~0822+3542 and I~Zw~18 by a factor of 2--4,
we describe below in more detail
various parameters of HS~0822+3542 and compare them to other well-known
young galaxy candidates.

\subsection{Chemical abundances}

The results from the chemical abundance determination are presented in
Table~\ref{t:Chem}. For the sake of comparison, we also show
the data for the other two low metallicity galaxies
SBS~0335--052 and I~Zw~18.
A two-zone photoionized H{\sc ii} region has been assumed; the electron
temperature for the high-ionization region has been
obtained from the [O{\sc iii}] $\lambda$\,4363/($\lambda$\,4959+5007) ratio
using a five-level atom model (Izotov \& Thuan \cite{IT99})
and for the low-ionization region by using the empirical relation
between both electron temperatures
(Izotov, Thuan \& Lipovetsky \cite{Izotov97b}). The electron density
was derived from the [S{\sc ii}]$\lambda$\,6717,6731 ratio.
Abundances have been calculated following
Izotov, Thuan \& Lipovetsky (\cite{Izotov94}, \cite{Izotov97b}).

The oxygen abundance in HS~0822+3542 is slightly higher than in the other 
two galaxies. This galaxy is, therefore, one more object filling the gap 
between the  metallicity of I~Zw~18 and those for the bulk of BCGs.
The comparison of the data on element abundances in HS~0822+3542
with the abundance ratios from Izotov \& Thuan (\cite{IT99}) shows that
all data for HS~0822+3542 agree with the derived average values 
for the sample of low-metallicity BCGs. The 
largest deviation occurs in the N/O ratio. But if we 
take into account the large uncertainty in the flux of [N{\sc ii}] $\lambda$6583
emission line blended with H$\alpha$, the N/O 
is consistent with the expected value for this kind of galaxies. 

We thus conclude that the measured ratios of heavy element abundances to
that of oxygen in HS~0822+3542 follow the same relation as in all metal-poor BCGs.
In particular, nitrogen is likely synthesized in massive stars that also
produce oxygen, neon and sulfur, as proposed by Izotov \& Thuan (\cite{IT99}).

We note, that low measured value of $C$(H$\beta$)=0.005$\pm$0.04,
transfered to a 2$\sigma$ upper limit: $C$(H$\beta$) $<$ 0.085, equivalent
to $E(B-V)$ = \mbox{0.68 $C$(H$\beta$)} $<$ 0.058, is consistent with
$E(B-V)$ = 0.047$\pm$0.007 from Schlegel et al. (\cite{Schlegel98}), which
was used to correct measured colours and absolute magnitude.
Anyway the net internal extinction A$_B$ in HS~0822+3542 seems to be very low,
implying a low dust content, what is also consistent with the object
extremely low metallicity.

\subsection{Morphology and colours}

The $R$-band image of the galaxy, with filamentary structures
at the periphery of the low surface brightness region, is typical of dwarfs
with strong SF activity.
Curiously, the appearance of HS~0822+3542 is morphologically similar, 
on the same angular scales,
to that of SBS~0335--052E (Melnick et al. \cite{Melnick92};
Thuan et al. \cite{TIL97}), including some arcs and filaments. The integrated
($B-V$) and ($V-R$) colours of HS~0822+3542  are very blue
($0\fm32$ and $0\fm17$, respectively), similar to those of SBS~0335--052E.

Below, we describe the surface brightness (SB) distribution of
HS~0822+3542. We show
in Fig.~\ref{fig:SBPs} the SB profiles in $B$,  $V$ and
$R$-bands, which look very similar, with some deviations in $R$-band
in the middle part of the profile. This is presumably due to additional H$\alpha$
emission from the filamentary structure (see Fig.~\ref{fig:HS_direct}).
The $B$-band SB profile (Fig.~\ref{fig:SBP})
indicates the presence of two components: the central bright compact
body and an exponential disc, dominating the light in
the outer part of the profile  (hereafter LSB component).
The parameters of both components, derived from the fitting of the SB
profiles in  $B$, $V$ and $R$, are given in Table~\ref{tab:struct_par}.

The corresponding exponential scale lengths are $\alpha_{\rm E}^B$ = 86$\pm$1 pc,
$\alpha_{\rm E}^V$ = 86$\pm$2 pc and $\alpha_{\rm E}^R$ = 84$\pm$2~pc.
A comparison with the same parameters derived for the LSB component of other BCGs
(Papaderos et al.~\cite{Papa96};  Papaderos et al.~\cite{Papa98};
Doublier et al.~\cite{Doublier99a})
shows that the scale length of HS~0822+3542 is smaller than that of
any BCG studied in the cited works.
Even the ``ultracompact'' dwarf galaxy POX~186 (Doublier et al.~\cite{Doublier99b})
has a scale length more than twice larger (180 pc).
However HS~0822+3542 is not the only extreme case in its small disc ``size''. 
At least two galaxies
have comparable or smaller scale lengths: GR~8
(Mateo~\cite{Mateo98}) and Tol~1116--325 (Telles et al.~\cite{Telles97}).

Integrating the SB profile of the underlying exponential disc we obtain its total 
$B$ magnitude:
\begin{equation}
B_{\rm disc} = \mu_{\rm E,0} - 5\log(\alpha_{\rm E}) - 1.995,
\end{equation}
where $\alpha_{\rm E}$ is in arcsec.

With the resulting $B_{\rm disc}$ = 18\fm22 we can estimate also the luminosity
of current SF burst and the corresponding brightening of the galaxy.
The brightening is quite modest, about 30\%. The light of current burst
corresponds to $B_{\rm burst}$ = 19\fm46 and a luminosity
of $M_{B} = -$11\fm1.

Since the scale lengths for the underlying disc in all three bands are identical 
within small uncertainties (1--1.5\%), we can accept as
a first approximation that the scale length of the disc is a unique one for
$B$, $V$ and $R$, equal to the average $<\alpha_{\rm E}>$ = 1.40$\pm$0.02
arcsec, and corresponding to 85$\pm$1 pc. Hence, the underlying disc
has no colour gradient, and its colours ($B-V$)$_{\rm disc}$
and ($V-R$)$_{\rm disc}$
can be approximated by the colours of its central SB
$$(B-V)_{\rm disc} = \mu_{\rm E,0}^B - \mu_{\rm E,0}^V  = 
0\fm10\pm0\fm06, $$
$$(V-R)_{\rm disc} = \mu_{\rm E,0}^V - \mu_{\rm E,0}^R  = 
0\fm29\pm0\fm09, $$
respectively.
The disc colours can also be obtained from its integral $B$, $V$ and $R$ magnitudes
(column 7 in Table~\ref{tab:struct_par}):
$$(B-V)_{\rm disc} = 0\fm12\pm0\fm10, $$
$$(V-R)_{\rm disc} = 0\fm22\pm0\fm16, $$
consistent with the estimates above. The uncertainties of the latter colours
are derived from the errors of the integrated disc magnitudes, from the
propagation of errors in equation (2). They are $0\fm07$, $0\fm08$
and $0\fm14$ for $B$, $V$ and $R$, respectively.

A close coincidence of the scale lengths of BCG LSB components in the different
filters was found by Papaderos et al.\, (\cite{Papa98}) for SBS~0335--052E.
The same correlation for disc scale lengths in different bands
was shown to exist for a sample of 19
BCGs (Doublier et al. \cite{Doublier99a}).
However, the colours $(B-R)_{\rm disc}$ for BCGs with small scale lengths
from Doublier et al. are redder and lie in the range of 0\fm61--1\fm47,
with an average $(B-R)_{\rm disc}$ = 1\fm11$\pm$0\fm30.
The most similar to HS~0822+3542 in this colour
(with $(B-R)_{\rm disc}$ = 0\fm61) is
 SBS~0940+544, with
12+log(O/H) = 7.43$\pm$0.01 (Izotov \& Thuan~\cite{IT99}).

In Fig.~\ref{fig:phot_colours} we show the distributions of observed
($B-V$) (filled squares) and ($V-R$) (open circles) colours, as functions
of the effective radius.
Mean
disc colours are shown by filled and open rectangles to the right of the
observed colour profiles. Note that the ($V-R$) colours are shifted  by 1\fm0 in
Fig.~\ref{fig:phot_colours}, to minimize confusion.
Note also that ($V-R$) is even bluer for the outermost part of disc
($R^* > $3\arcsec): 0\fm16$\pm$0\fm12.

Effective sizes for the central gaussian component in the different bands 
also coincide, within small
uncertainties. We can therefore think of this component as a
body with a homogeneous colour distribution.
We note also, that this compact gaussian component is significantly larger
than we could expect for a point-like source convolved with the point
spread function
(PSF). The FWHM for stellar images measured on the  CCD frames near
the galaxy is  1\farcs25, whereas it is 
$\alpha_{\rm G}$=1\farcs6 (Table~\ref{tab:struct_par}) for the bright
central region. 
The deconvolved FWHM for this component amounts to $\approx$ 1\farcs0.
Comparing with the structure of star-forming regions in other
BCGs, it is natural to assign the enhanced
optical emission in the central part of HS~0822+3542 to a young massive-star 
cluster formed in the current SF burst and its associated  H{\sc ii}
region. The characteristic linear radius of this complex
is $\sim$ 30 pc, comparable to the size of the star cluster
R136 in LMC  (Walborn \cite{Walborn91}). Accordingly, the absolute
$B$ magnitude of this bright component --11\fm1 is near the lower limit of the
range found with HST for super-star clusters by O'Connel et al.
(\cite{O'Connel94}) in two nearby star-bursting galaxies.

The total size of HS~0822+3542 out to the $\mu_{B}$=25 mag arcsec$^{-2}$ 
isophote can be approximated by an ellipse with major
and minor axes 14\farcs8 and 7\farcs4, or 900 by 450 pc, respectively.

As we have shown above, the SB distribution of HS~0822+3542 can be fitted
by two components. While the brighter and more compact central component
seemingly corresponds to the complex of young  massive stars formed during the
current SF burst and its associated H{\sc ii} region, the underlying exponential
disc can consist of older stars. The latter could be formed either
during a previous much earlier SF episode, or relatively recently, in a
precursor of the current SF burst,
which could be significantly displaced ($\sim$150--200 pc) from the
underlying disc centre. If the current SF burst and the previous SF
activity are causally connected, the propagation time of the SF wave from the
underlying disc centre, with a typical velocity of 10~km s$^{-1}$ (e.g.,
Zenina et al. \cite{Zenina97}) would be only 15--20 Myr.
In order to check this option one could search for the He{\sc i}
absorption features of early B stars in the underlying disc.

To check for emission from  older stellar
populations, which may reside in the region outside the current SF burst,
we compare the colours of underlying disc with those predicted for
various models, as well as with similar parameters for  other young
galaxy candidates.

The very blue colours of the disk, after correcting for the extinction
in the Galaxy
($B-V$)$_0$ = $0\fm05\pm0\fm06$
and ($V-R$)$_0$ = $0\fm26\pm0\fm09$
are reasonably consistent with the predictions for an instantaneous SF
burst with a Salpeter IMF, a metallicity of 1/20 $Z_{\odot}$ and an age
of $\sim$ 100 Myr (Leitherer et al. \cite{Leitherer99}).
If prolonged star formation is assumed, then the age of the oldest 
stellar population in the extended disc can be larger than that of the 
instantaneous burst. The observed colours can also be explained by
continuous SF with a constant star formation rate from 500 Myr to 20
Myr ago. However, older stellar populations with an age of
10 Gyr are excluded. The derived age is somewhat lower for the outermost 
part of the of the disc with a 
bluer ($V-R$) colour in comparison to its average value.
Integrated colours of ionized gas  are about ($B-R$) = 0\fm6--0\fm8
(Izotov et al.~\cite{Izotov97a}), what is redder than our observed values
in the underlying disc. Therefore if some gas emission inputs to the
integrated disc colours, then the true colours of stellar component should be
even bluer than measured.

These colours are very similar to those of the underlying nebulosity
in SBS~0335--052. The latter colours
are shown to be well explained by the radiation of ionized gas and A-stars
with the ages of no more than 100 Myr, created in the current SF episode.
This SF episode is prolonged and may represent a
propagating SF wave (Papaderos et al. \cite{Papa98}), similar to what is
suggested for another candidate young galaxies I~Zw~18 (Izotov et al.
\cite{Izotov2000}) and CG~389 (Thuan et al. \cite{TIF99}).
This similarity of LSB disc colours of HS~0822+3542 and
SBS~0335--052E suggests that up to the radial distances of $\sim$300 pc
the input of stellar population with the ages larger than 100 Myr to the
disc radiation is undetectable.

As a first approximation, the colours of the underlying disc do not
contradict the possibility that the current burst is the first SF episode
in this galaxy.
However since the emission of ionized gas can add significantly
to the total radiation from the volume within 300--400 pc, 
further photometric and spectroscopic observations of HS 0822+3542 are required
to account properly for the contribution of gaseous emission.

\subsection{Current star formation rate}

The star formation rate (SFR) of the current SF episode can be estimated
from the total H$\alpha$ luminosity. The H$\alpha$ flux was measured within
the slit, and was corrected for the missing light outside the slit.
A correction factor of 2.33 was calculated, using the brightness
profile along the slit and assuming circular symmetry.
For the derived total flux $F$(H$\alpha$)=5.8$\times10^{-14}$\
erg\ s$^{-1}$cm$^{-2}$ we obtain a total H$\alpha$ luminosity of 10$^{39}$\
erg\ s$^{-1}$.

This H$\alpha$ luminosity corresponds to a current SFR
of $\sim$ 0.007 $M_{\odot}$ yr$^{-1}$ (Hunter \& Gallagher \cite{Hunter86}) 
assuming a Salpeter IMF with a  0.1 $M_{\odot}$ lower mass cutoff.
The gaseous mass converted into the stars during a 
3 Myr burst is  $\sim$ 2$\times$10$^{4}$ $M_{\odot}$. 
The unknown contribution of the gaseous emission to the light from the LSB
component further complicates the stellar mass estimate. If only stars with age
$\le$ 100 Myr contribute to
the LSB luminosity, then the total mass of stars in the underlying
disc with $M_{B,{\rm disc}}$=--12\fm3
is 1.3$\times$10$^6$ $M_{\odot}$ (from e.g., Leitherer et al. 
\cite{Leitherer99}). This is much smaller than the total 
neutral gas mass of $\approx$3.0$\times$10$^{7}$ $M_{\odot}$, 
accounting for a helium mass fraction of 0.25.

The total mass of ionized gas can
be obtained from the average mass density inside the  H{\sc ii}
region and its volume. An average mass density corresponding to the
average electron density $N_{\rm e}$ of 1 cm$^{-3}$ within a volume
with a diameter of 0.5 kpc yields a total ionized gas mass of 10$^{6}$
$M_{\odot}$.

The above estimates show that baryonic matter in HS 0822+3542 is dominated by
the gaseous component.

\subsection{{\rm H}{\sc i} and dynamical mass }

The integrated H{\sc i} line flux, the characteristic widths of 21 cm line profile $W_{50}$
and $W_{20}$ (for a Hanning smoothing of 10.6~km s$^{-1}$), and the derived H{\sc i}
mass $M_{\rm HI}$ for HS~0822+3542 are presented in Table~\ref{tab:Main_par}.
The small H{\sc i} mass 2.4$\times10^{7}$$M_{\odot}$  of this galaxy
is consistent with its low optical luminosity.
The very narrow H{\sc i} profile is indicative of its very low amplitude of rotational
velocity, which does not exceed 30~km s$^{-1}$. It is difficult to assess the
inclination angle correction, since the optical morphology can be
unrelated to global properties of the associated H{\sc i} cloud, as exemplified by the
case of SBS~0335--052 (Pustilnik et al. \cite{Pus2000}).
The role of chaotic gas motions is, in general, more important in very low mass
galaxies, where the amplitude of random velocities reaching ten km s$^{-1}$ or more
can be commensurate with the rotational velocity.

The measured H{\sc i} mass and the profile width are in the range characteristic of very
low mass galaxies.  The mass-to-light parameter $M$(H{\sc i})/$L_{B}$ =
1.40 $M_{\odot}$/$L_{\odot}$,
is comparable to those of I~Zw~18 and SBS~0335--052E.
It is not as high as in some gas-rich dwarfs from van Zee et
al. (\cite{vanZee97}), but these galaxies, although they
have a few H{\sc ii} regions, are in relatively quiescent state.
The extremely metal-poor BCGs discussed here experience  significant
luminosity enhancement due to very intense current  and recent
SF activity; this results in a significant decrease of their mass-to-light
ratios relative to their non-active state.

A rough estimation can be made also on the dynamical mass of HS~0822+3542.
From the width of H{\sc i} profile at the 20\% level,
a maximum rotation velocity of 30~km s$^{-1}$ can be assumed. 
The extent of the H{\sc i} cloud associated with a BCG is normally many times
larger than its optical size.
The optical radius $R_{25}$ is the radius
of a disc galaxy at the isophotal level $\mu_{B}$ = 25.0 mag arcsec$^{-2}$.
A conservative lower limit to the ratio of H{\sc i}-to-optical radii
is 4 (see, e.g. Taylor et al.~\cite{Taylor95};
Chengalur et al.~\cite{Chengalur95}; Salzer et al.~\cite{Salzer91};
van Zee et al.~\cite{vanZee98}; Pustilnik et al. \cite{Pus2000}).
By equating gravitation and centrifugal force at the edge of H{\sc i} disc
an estimate of the dynamical mass inside 1.5 kpc is obtained; this is
3.4$\times$10$^{8}$ $M_{\odot}$,  
one order of magnitude larger than the total visible mass of the
galaxy
$M_{\rm neutral}$ + $M_{\rm stars}$ + $M_{\rm HII~region}$. 

Even rather unprobable case of H{\sc i} radius equal to the optical one
results in the total dynamical mass $\sim$3 times higher than the visible mass.
Hence, HS~0822+3542 like other extremely metal-deficient dwarf 
galaxies, is dynamically dominated by a dark matter halo, supporting the
modern view of primeval galaxy formation process (e.g. Rees \cite{Rees88}).
In turn the mass of its DM halo  is one of the smallest for galaxies.

\section{A candidate young galaxy?}

The properties of HS~0822+3542 presented above (extremely
low abundance of heavy elements, very blue colour of underlying nebulosity,
and extremely small mass ratio of stars to neutral gas) 
suggest that this could be the nearest candidate young dwarf galaxy 
forming its first stellar generation. However, 
on the basis of the present data, we cannot exclude the presence of an
underlying older stellar populations originating from earlier SF episodes.

Similar studies of the two previously known young galaxy
candidates I~Zw~18 and the pair SBS~0335--052E/0335--052W (Izotov et al.
\cite{Izotov97a}; Thuan \& Izotov \cite{TI97}; 
Pustilnik et al. \cite{Pus97}; Lipovetsky et al.
\cite{Lipovetsky99}) have not been conclusive whether such young systems
really do exist.

While the chemical properties (see Table~\ref{t:Chem})
of these candidates seem
to follow a general trend and behave quite homogeneously, other global 
properties have still to be understood.
From Table~\ref{tab:Main_par} it appears that they cover a broad range of
neutral hydrogen masses and  blue luminosities (a factor of $\sim$ 40--60).
The same is true for their current SFRs and for the mass of stars formed in a single
star formation episode. 

We suggest that a simple  linear scaling between
several important parameters  holds for forming galaxies, at least
in the range of baryon masses of (0.3--20)$\times$10$^{8}$ $M_{\odot}$ (see
Table 1, and assuming that most of baryons in these BCGs are in atomic hydrogen
and helium). This is important both for the analysis of conditions capable of
maintaining  pristine gas clouds stable for $\sim$ a Hubble time,
and for the  planning of further searches for such objects. In particular,
such low-mass objects 
can represent a significant fraction of the Ly$\alpha$ absorbers
at high redshifts.

One more indirect argument for the possible youth of the
three BCGs considered here comes from the calculations of mass and heavy
elements loss in dwarfs with active SF (Mac Low \& Ferrara \cite{MacLow99}).
These show that the rate of metal loss is strongly dependent on the baryon
mass in the range of 10$^7$ to 10$^9$ $M_{\odot}$. Therefore, if these
extremely metal-poor galaxies with masses ranging from  3$\times$10$^7$ to
2.5$\times$10$^9$ $M_{\odot}$ were not in their first SF episode, we should
expect significant differences in their observational properties. For the
three BCGs discussed here we do not find these differences.

From the surface density of already known objects with
extremely low metallicity and small radial velocity one can expect to find
at least ten more such galaxies within 15 Mpc, if the search will be extended
to the entire sky.

\section{Conclusions}

From the data and discussion above we reach the following conclusions:
\begin{enumerate}
\item HS~0822+3542 is a new nearby ($D$ = 12.5 Mpc)  galaxy
  with oxygen abundance 12 + log(O/H) = 7.35.
  After I~Zw~18 and SBS~0335--052 this is the third lowest
  metallicity  object among Blue Compact Galaxies.
\item Its very low metallicity, very small stellar mass fraction (0.05 
   relative
   to the entire baryon mass) and blue colours of the underlying disc
   [($B-V$)$_0$ = 0\fm05, ($V-R$)$_0$ = 0\fm26]
imply that
   this  is one of the few candidates to be a local young galaxy,
   forming its first generation of stars.
\item HS~0822+3542 is 50--60 times less luminous and massive than another
   candidate young galaxy SBS~0335--052. This implies a broad range
   of global parameters for the candidate young galaxies. A linear
   scaling between several important parameters of such galaxies probably exists,
   including  parameters related to the SF burst.
\item The dynamical mass estimate using the width of the H{\sc i} profile
   and a typical H{\sc i} gas extent relative to the optical size, leads to the
   conclusion that HS~0822+3542 is dynamically dominated by a dark matter halo.
\item Higher S/N long-slit spectra than presented here, and deep H$\alpha$ images
 are needed to follow  the ionized gas extent.
   Resolved H{\sc i} maps will be very helpful to study
   the dynamics of its ISM and the parameters of its DM halo.
\end{enumerate}

\begin{acknowledgements}

The authors thank A.I.Kopylov for $BVR$ calibration frames.
This work was partly supported by the INTAS grant 96-0500. The SAO authors
acknowledge partial support from the Russian Foundation for Basic Researches
by grant No. 97-2-16755 and the Center for Cosmoparticle Physics ``Cosmion''.
Part of the data presented here have been taken using ALFOSC, which is
owned by the Instituto de Astrofisica de Andalucia (IAA) and operated at
the Nordic Optical Telescope under agreement between IAA and the NBIfA of
the Astronomical Observatory of Copenhagen.
One of us (AK) acknowledges the support by the Junta de Andalucia
during his stay at the IAA.
The authors thank the anonymous referee for useful suggestions which allowed
to improve the presentation of several points.
The authors acknowledge the use of the NASA/IPAC Extragalactic Database
(NED) and Lyon-Meudon Extragalactic Database (LEDA).

\end{acknowledgements}

\end{document}